\documentclass[floats,floatfix,showpacs,amssymb,prd,twocolumn,superscriptaddress,nofootinbib]{revtex4-1}

\usepackage{anyfontsize}
\usepackage{amssymb}
\usepackage{amsmath}
\usepackage{amsfonts}
\usepackage{amsbsy}

\usepackage{newtxtext}
\usepackage{newtxmath}

\usepackage{tensor}
\usepackage{mathrsfs}

\usepackage{bm}

\usepackage[utf8]{inputenc}

\usepackage{graphicx}
\usepackage{epsfig}
\usepackage{epstopdf}

\usepackage[usenames,dvipsnames]{xcolor}

\usepackage{verbatim,mathtools,needspace,enumitem,etoolbox}

\definecolor{linkcolor}{rgb}{0.0,0.3,0.5}

\usepackage[unicode, colorlinks=true, linkcolor=linkcolor, citecolor=linkcolor, filecolor=linkcolor,urlcolor=linkcolor, pdfusetitle]{hyperref}

\usepackage{epstopdf}

\usepackage{bm}
\usepackage{dcolumn}

\usepackage{longtable}
\usepackage{enumerate}
 \usepackage[normalem]{ulem}

\definecolor{romared}{RGB}{142,0,28}
\hypersetup{colorlinks=true,
            citecolor=romared,
            linkcolor=romared,
            urlcolor=romared}
\newcommand{\dd}{{\rm d}}

\newcommand{\be}{\begin{equation}}
\newcommand{\ee}{\end{equation}}
\newcommand{\op}{\left}
\newcommand{\cl}{\right}

\newcommand{\vp}{\varphi}

\newcommand{\mpl}{M_{\rm Pl}}

\begin{document}
\title{Self-interactions and Spontaneous Black Hole Scalarization}

\author{Caio F.~B.~Macedo}
\email{caiomacedo@ufpa.br}
\affiliation{Campus Salin\'opolis, Universidade Federal do Par\'a,
Salin\'opolis, Par\'a, 68721-000, Brazil}

\author{Jeremy Sakstein}
\email{sakstein@physics.upenn.edu}
\affiliation{Center for Particle Cosmology,
Department of Physics and Astronomy,
University of Pennsylvania,
209 S. 33rd St., Philadelphia, PA 19104, USA}

\author{Emanuele Berti}
\email{berti@jhu.edu}
\affiliation{Department of Physics and Astronomy,
Johns Hopkins University, 3400 N. Charles Street, Baltimore, MD 21218, USA}

\author{Leonardo Gualtieri}
\email{leonardo.gualtieri@roma1.infn.it}
\affiliation{Dipartimento di Fisica ``Sapienza''
Universit\`a di Roma \& Sezione INFN Roma1,
Piazzale Aldo Moro 5, 00185, Roma, Italy}

\author{Hector O. Silva}
\email{hector.okadadasilva@montana.edu}
\affiliation{eXtreme Gravity Institute,
Department of Physics, Montana State University, Bozeman, MT 59717 USA}

\author{Thomas P. Sotiriou}
\email{thomas.sotiriou@nottingham.ac.uk}
\affiliation{School of Mathematical Sciences,
University of Nottingham, University Park, Nottingham, NG7 2RD, UK}
\affiliation{School of Physics and Astronomy,
University of Nottingham, University Park, Nottingham, NG7 2RD, UK}

\raggedbottom

\begin{abstract}
{It has recently been shown that   nontrivial
  couplings between a scalar and the Gauss-Bonnet invariant can give rise to black hole spontaneous scalarization. Theories that exhibit this phenomenon are among the leading candidates for
  testing gravity with upcoming black hole observations.}
 All models considered so far have
  focused on specific forms for the coupling, neglecting scalar self-interactions. In this work, we take the first steps
  towards placing this phenomenon on a more robust theoretical footing by considering the leading-order scalar
  self-interactions as well as the scalar-Gauss-Bonnet coupling. {Our approach is consistent with the principles of  effective field theory and yields 
  the simplest and most natural model.} We find that a mass term for the scalar alters the
  threshold for the onset of scalarization, and we study the mass range over which scalarized black hole solutions
  exist. We also demonstrate that the quartic self-coupling is sufficient to produce scalarized solutions that are
  stable against radial perturbations, without the need to resort to higher-order terms in the Gauss-Bonnet coupling
  function. Our model therefore represents a canonical model that can be studied further, with the ultimate aim of
  developing falsifiable tests of black hole scalarization.
\end{abstract}

\date{\today}

\maketitle

\section{Introduction}
\label{sec:int}

The era of gravitational-wave observations has arrived. For the first time we can see the universe in gravitational
waves as well as optically, and this new window affords us the opportunity to test gravity in extreme spacetimes for the
first time. The LIGO/Virgo collaboration has already detected ten black hole (BH) mergers
\cite{Abbott:2016blz,LIGOScientific:2018mvr} and one neutron star merger \cite{Abbott:2017oio}. The latter has proved
incredibly powerful for testing and constraining infrared modifications of
gravity~\cite{Sakstein:2017xjx,Ezquiaga:2017ekz,Creminelli:2017sry,Baker:2017hug,Dima:2017pwp,Crisostomi:2017lbg,Langlois:2017dyl} (if the modifications are important for the late-time cosmology \cite{Franchini:2019npi}),
but ultraviolet (UV) modifications are more difficult to test. This is partly due to the numerical and theoretical
challenges that arise when extending computations of merger events to theories beyond general relativity (GR), but also
due to a theoretical roadblock: the no-hair theorems~\cite{Hawking:1972qk,Sotiriou:2011dz,Hui:2012qt} (see
e.g.~\cite{Berti:2015itd,Sotiriou:2015lxa,Herdeiro:2015waa,Sotiriou:2015pka,Cardoso:2016ryw} for reviews).  These preclude the existence of nontrivial
scalar hair (or scalar charges) for BHs, and so the dynamics of theories including new scalar degrees of freedom
(i.e. scalar-tensor theories) is similar to GR. One possible way forward is to instead use neutron stars as probes of UV
modifications of GR~\cite{Yagi:2013awa,Pappas:2014gca,Berti:2015itd,Pappas:2015npa,Yagi:2016ejg,Sakstein:2016oel,Babichev:2016jom,Doneva:2017jop,Sakstein:2018fwz,Pappas:2018csu}.  These are far more complicated objects since the equation of
state for nuclear matter is presently unknown, and, unlike BHs, neutron stars have higher-order multipole moments that
give rise to strong tidal effects. On the observational side, LIGO/Virgo has observed more BH mergers than neutron star
mergers \cite{LIGOScientific:2018mvr}, and this may well remain the case, even as more gravitational-wave detectors come
online and the existing ones are upgraded to improved sensitivities.

\subsection{Black hole spontaneous scalarization and effective field theory}

The considerations above have motivated a theoretical effort to find UV-modifications of GR that can circumvent the
no-hair theorems by violating some of their assumptions. Some of these theories exhibit \emph{spontaneous BH scalarization}~\cite{Doneva:2017bvd,Silva:2017uqg}, a phenomenon where
both the GR BH solution and novel BH solutions with scalar hair can exist. The phenomenon has been predicted for
static~\cite{Doneva:2017bvd,Silva:2017uqg} and, more recently, charged~\cite{Doneva:2018rou,Brihaye:2019kvj} BHs. 
This allows for the possibility that, even if all LIGO/Virgo detections to date have been compatible with GR, future
detections could be consistent with scalarized BH solutions.

The fundamental interaction responsible for
scalarization is the coupling between a scalar field $\phi$ and the Gauss-Bonnet invariant ${\cal
  G}=R^2-4R_{ab}R^{ab}+R_{abcd}R^{abcd}$, so that the action has the form
\begin{equation}
S=\frac{1}{2}\int\dd^4 x\sqrt{-g}\left[\frac{R}{8\pi G}-\frac{1}{2}\nabla_a\phi\nabla^a\phi+f(\phi)\mathcal{G}\right],
\label{eq:action0}
\end{equation}
where we are using units where $\hbar=c=1$, so that the Planck mass
$\mpl=(8\pi G)^{-1/2}$. In subsequent sections we will switch to units where $8\pi G=c=1$, which is
more suited to (and more common in) the study of BH solutions. In
Eq.~\eqref{eq:action0} we have chosen the same normalization and
conventions as in Ref.~\cite{Kanti:1995vq} (modulo an overall sign in
the definition of the Riemann tensor) and in
Refs.~\cite{Maselli:2015tta,Silva:2017uqg,Minamitsuji:2018xde,Silva:2018qhn},
while the scalar field $\phi^{\rm DY}$ in
Refs.~\cite{Doneva:2017bvd,Blazquez-Salcedo:2018jnn} has a different
normalization: $\phi^{\rm DY}=\phi/2$. A canonically normalized scalar
field $\phi^{\rm can}$ is such that $\phi^{\rm can}=\phi/\sqrt{2}$.

Reference \cite{Silva:2017uqg} proved a no-hair theorem for scalar-Gauss-Bonnet theories under certain conditions. Scalarization may occur when these conditions are violated.
The essential requirement is that the coupling function $f(\phi)$ has at least one stationary point at
some $\phi=\bar\phi$ such that $f(\bar\phi)=0$. GR BHs correspond to solutions
with $\phi=\bar\phi$, but this configuration may be unstable for certain BH masses or model parameters. When this is the
case, the field rolls away, and the BH acquires scalar hair. Apart from this requirement, there is no guiding principle
for choosing $f(\phi)$. The patent choice $f(\phi)=\phi^2/2\mathcal{M}^2$~\cite{Silva:2017uqg} (where $\mathcal{M}$ is a new mass scale) produces
scalarized BHs that are unstable to radial perturbations \cite{Blazquez-Salcedo:2018jnn}. This can be resolved by
including higher-order terms, in particular $f(\phi)=\phi^2/2\mathcal{M}^2+c\phi^4/{\mathcal{M}}^4$
\cite{Minamitsuji:2018xde,Silva:2018qhn}, or by assuming a more complicated function: for example, exponential couplings
$f(\phi)=\exp(\beta\phi^2/2\mathcal{M}^{2})$ with both positive and negative signs for $\beta$ have been considered in
the literature~\cite{Doneva:2017bvd,Antoniou:2017acq}.

These solutions are somewhat unsatisfactory from a theoretical perspective. Since we lack a UV-completion for these
models, it would be more appropriate to construct the theory using the principles of effective field theory
(EFT)~\cite{Georgi:1994qn,Donoghue:1994dn,Burgess:2007pt}. From this perspective, relying on higher-order corrections to the coupling function in order to stabilize the BH solutions implies that higher-dimensional
operators are competing with (supposedly leading) lower-dimensional operators. This suggests that  operators that
have been omitted can be  just as important, and therefore these solutions are outside the range of validity of the
EFT. Moreover, without any enhanced symmetry protecting the form of special choices of the coupling functions (and the action in general), it
is  likely that these theories are radiatively unstable. {We note that there is currently no known enhanced symmetry of the exponential or quartic couplings, though this is by no means a proof that there cannot be one. Similarly, it is possible that such couplings arise as a truncation of a UV-complete theory and just appear fine-tuned from an IR perspective~\cite{Heckman:2019dsj}. }

In the coming decade and beyond, LIGO/Virgo will be upgraded to higher sensitivities and additional detectors will come
online. Hundreds or thousands of detections are anticipated, and it therefore behooves us to make theoretical
predictions from robust models that are stable from a QFT point of view. The
main purpose of this paper is to take a first step
towards placing the phenomenon of BH spontaneous scalarization on a more robust theoretical foundation by
constructing the theory using EFT principles.

When viewed as an EFT, spontaneous scalarization is a 
phenomenon occurring in theories where a $\mathbb{Z}_2$-symmetric scalar (i.e. the action is invariant under $\phi\rightarrow -\phi$) is coupled to a
massless spin-2 particle. We should therefore build our action out of operators that are invariant under this
symmetry. In particular, the leading-order (relevant and marginal) operators are not Gauss-Bonnet couplings, but rather
a mass term and a quartic self-interaction\footnote{The leading-order scalar-graviton coupling is $\phi^2R$, but this does not lead to BH scalarization, so we will not include it in this work. This operator does not contribute to the scalar?s equation of motion on a Ricci-flat GR solution, which  means it cannot alter the threshold for the onset of BH scalarization. Note however that it can contribute to the effective mass on a scalarized BH or a neutron star background.}. One should
supplement these with irrelevant operators suppressed by some cut-off scale $\mathcal{M}$, which will include a
quadratic scalar-Gauss-Bonnet coupling at lowest order. For this reason, we will mainly study the action
\begin{align}\label{eq:actionmin}
  S=&\frac{1}{2}\int\dd^4 x\sqrt{-g}\left[\mpl^2R-\frac{1}{2}\nabla_a\phi\nabla^a\phi\right.\\
    &\left.-\frac{1}{2}\mu^2\phi^2-\frac{1}{2}\lambda\phi^4+\frac{\phi^2}{2\mathcal{M}^2}\mathcal{G}\right].\nonumber
\end{align}

Later on, we will also include the quartic scalar-Gauss-Bonnet
coupling in order to provide a different and well-studied stable model
against which we can compare the effects of the
self-interactions. This coupling is higher order and it will not
give the leading order effect from an EFT perspective, but it is
included for the purpose of comparison with previous
studies~\cite{Doneva:2017bvd,Silva:2017uqg,Blazquez-Salcedo:2018jnn,Silva:2018qhn}.
Boson star and BH solutions have recently been studied in a similar class of theories~\cite{Baibhav:2016fot,Brihaye:2018grv}. Scalar-tensor theories with a self-interacting potential, but without a Gauss-Bonnet term, were considered
in~\cite{Cheong:2018gzn,Arapoglu:2019mun,Staykov:2018hhc}.

\begin{figure*}[t]
\includegraphics[width=\columnwidth]{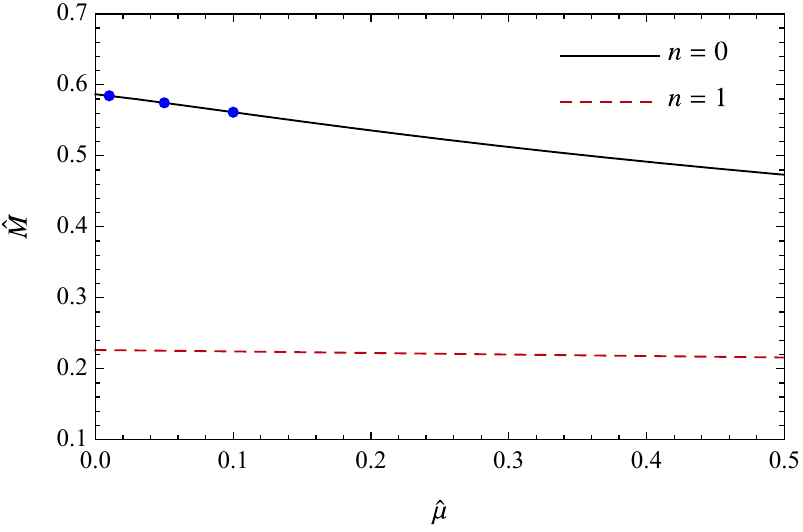}
\includegraphics[width=\columnwidth]{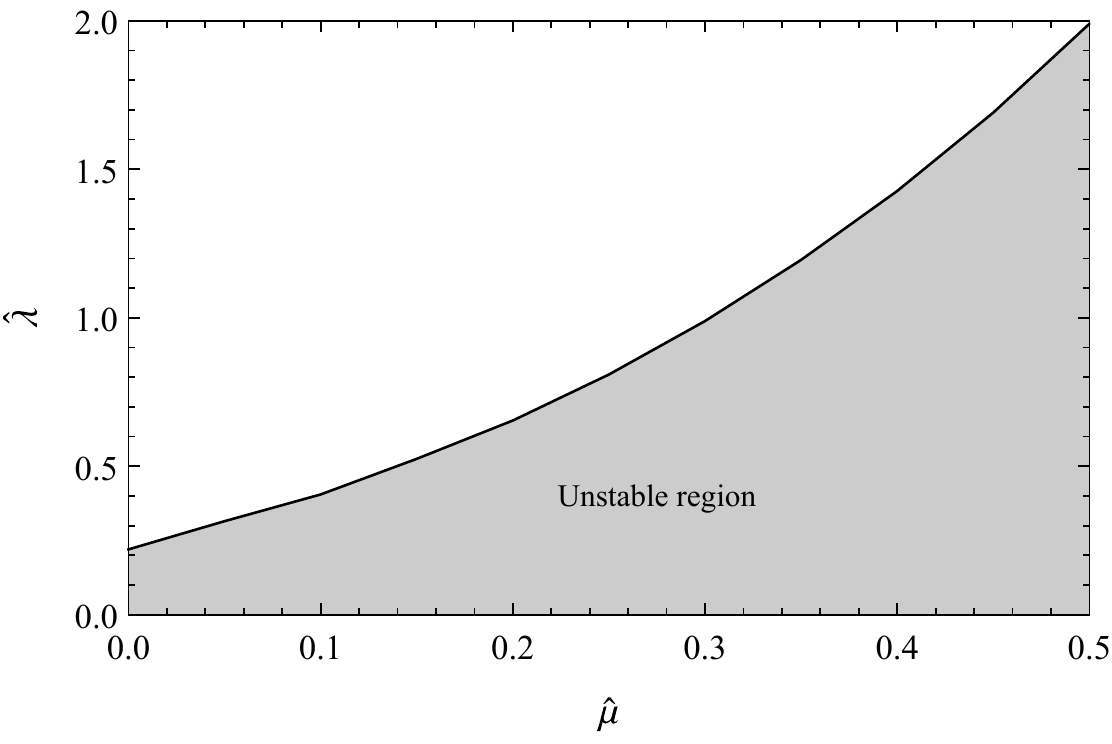}
\caption{\textbf{Left}: Critical threshold for scalarization as function
  of dimensionless mass $\hat\mu$ for the scalarized
  solutions with $n=0$ and $n=1$ nodes in the scalar profile. {Scalarized solutions with $n>0$ are unstable under radial
  perturbations. The scalarization threshold is
  independent of $\hat \lambda$, and stable scalarized solutions may exist for masses $\hat{M}$ below the lines (for $n=0$).}
  Conversely, the Schwarzschild solution is stable for masses
  larger than the $n=0$ threshold and unstable for masses below this value.
  Blue dots mark the values of $\hat \mu$ studied more in detail in Fig.~\ref{fig:sch_ins}.  \textbf{Right}: Phase
  diagram for stable, scalarized BH solutions in the $(\hat \mu,\,\hat \lambda)$ plane. All scalarized BH solutions in
  the gray region are radially unstable.}
\label{fig:scalarized_sols}
\end{figure*}

Of course, there are other operators that one could write down, such as a term $\propto \phi^6$ in the potential or
scalar-curvature couplings such as $\phi^2R$, but in this work we will restrict our focus to understanding how
scalarization works when only the leading-order operators (including the leading-order scalar-Gauss-Bonnet coupling) are
included, since this is the minimal input required to produce the phenomenon. We postpone the more arduous task of
determining the unique set of dimension-six operators that contribute to this theory, and a full explorative study of
the resultant parameter space, for future work.

\subsection{Executive summary}

In this article we study the existence, stability, and properties of scalarized BHs in the theory defined by the action
\eqref{eq:actionmin}, including the subtleties and conceptual issues that arise due to the inclusion of a mass for the
scalar.

We find (as previously noted in~\cite{Brihaye:2018grv}) that including a mass term for the scalar alters the threshold
for the onset of scalarization. Most notably, we find that the quartic self-interaction is sufficient to stabilize some
scalarized BHs, and higher-order scalar-Gauss-Bonnet couplings are not required. For this reason, and because the theory is a robust EFT, the action
\eqref{eq:actionmin} represents the leading canonical model with which to study spontaneous BH scalarization.

The action in Eq.~\eqref{eq:actionmin} uses units where $\hbar=c=1$, which are useful for understanding the theory from
an EFT perspective. For the purposes of calculating, it is more convenient to use geometrized units where $8\pi
G=c=1$. Furthermore, we will rescale the field so that $\phi$ is dimensionless by defining (before the change of units)
$\phi=\mpl\varphi $. In the new units, the action~\eqref{eq:actionmin} reads
\begin{equation}
S = \frac{1}{2}
\int {\rm d}^4x
\sqrt{-g}
\left[
R - \frac{1}{2}\nabla_a\varphi\nabla^a\varphi - V(\varphi) + f(\varphi){\cal G}
\right]\,,
\label{eq:action}
\end{equation}
where the potential is
\begin{equation}
  V(\varphi)=\frac{1}{2}\mu^2\varphi^2+\frac{1}{2}\lambda\varphi^4\,,
\label{eq:potential}
\end{equation}
the coupling function is
\begin{equation}
f(\varphi)=\frac{1}{8} \eta \varphi^2\,,
\label{eq:coupling_fun}
\end{equation}
and $\eta$ has units of $[\textrm{Length}]^2$. Note that the parameters $\mu$ and $\lambda$ appearing in the potential \eqref{eq:potential} have units of
$[\textrm{Length}]^{-1}$ and $[\textrm{Length}]^{-2}$, respectively. In order to compare our results with those of~\cite{Silva:2018qhn}, in part of our analysis we will also include a
quartic term in the coupling function, i.e. $f(\varphi)=(\eta\varphi^2+\zeta\varphi^4)/8$, where $\zeta$ has units of
$[\textrm{Length}]^{2}$ {(see Appendix~\ref{app:quartic})}. Finally, note that in this work we will only consider $\mu^2>0$. One could consider $\mu^2<0$,
which would give a global minimum of the potential at some $\vp\ne0$. This sign choice would require the addition of a
cosmological constant to cancel the net vacuum energy at the new minimum in order for the theory to admit
asymptotically-flat spacetimes. Since the aim of this work is to discern the effects of the scalar self-interactions on
the canonical model of spontaneous scalarization, we prefer not to include this more technical, and quantitatively
different possibility.

\begin{figure*}
\includegraphics[width=\columnwidth]{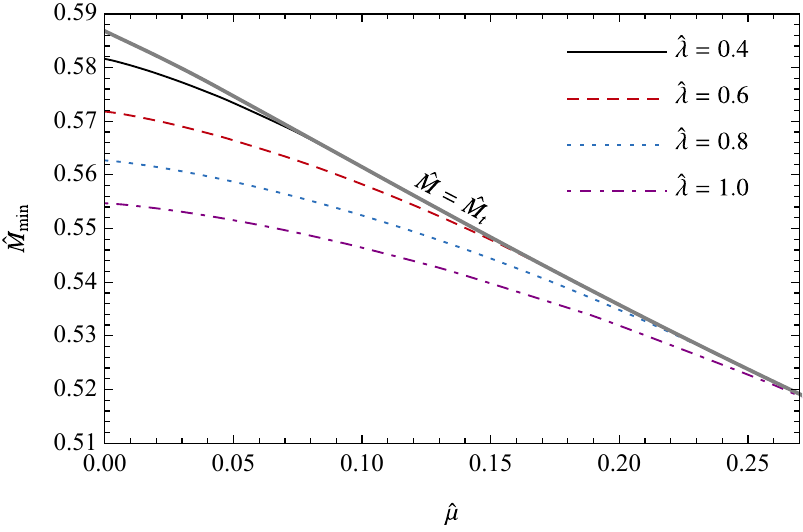}
\includegraphics[width=\columnwidth]{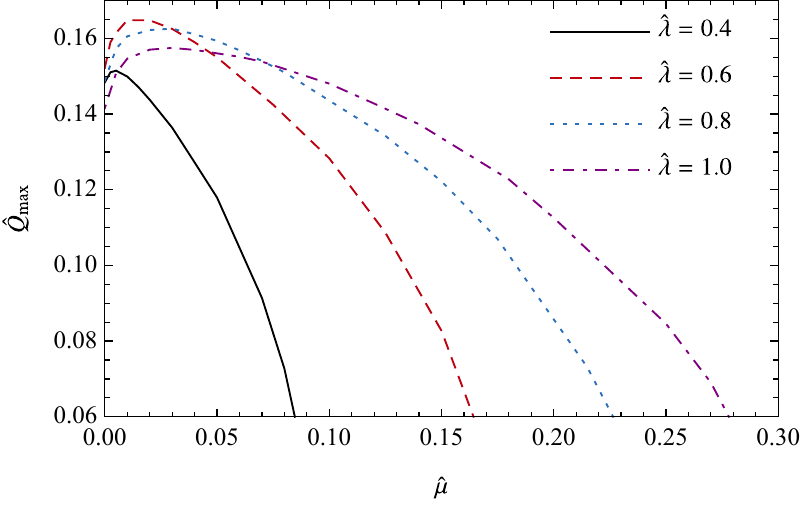}
\caption{Mass and charge of the marginally stable scalarized BH configurations corresponding to the blue dot configurations displayed in
  Fig.~\ref{fig:scalarized}. These represents the minimum mass and maximum charged that stable scalarized BHs can
  have.}
\label{fig:MQ}
\end{figure*}

With these conventions in place, we now summarize the main results of this work. First, we introduce the dimensionless
mass and scalar charge of the solutions
\begin{align}
  \hat{M}\equiv M/\eta^{1/2}, \quad
  \hat{Q}\equiv Q/\eta^{1/2},
\end{align}
as well as the dimensionless coupling parameters
\begin{align}
	\hat\mu\equiv \mu\eta^{1/2}, \quad
	\hat\lambda\equiv \lambda\eta, \quad
	\hat\zeta\equiv \zeta/\eta.
\end{align}
For the purpose of understanding the changes introduced by the scalar potential, it will be useful to introduce an
effective potential, which is spacetime dependent:
\begin{equation}\label{eq:effpot}
V_{\rm eff}(\varphi)=\frac12\left(\mu^2-\frac\eta4\mathcal{G}\right)\varphi^2+\frac{1}{2}\lambda\varphi^4
\end{equation}
The equation of motion for the scalar is $\Box\varphi=V_{\rm eff,\,\varphi}(\varphi)$. In particular, there is an effective mass
for the scalar about the point $\varphi=0$
\begin{equation}\label{eq:effmass}
m_{\rm eff}^2=\mu^2-\frac\eta4\mathcal{G};\quad \hat m_{\rm eff}=m_{\rm eff}/\eta^{1/2},
\end{equation}
where we have defined a dimensionless effective mass for later convenience. Close to the BH, the contribution of the Gauss-Bonnet invariant dominates and the Schwarzschild solution is unstable due to a tachyonic instability (recall that $\mathcal{G}\sim M^2/r^6$ for the Schwarzschild metric). This gives rise to
spontaneous scalarization. Further away, the Gauss-Bonnet contribution is negligible, and the effective mass is
positive. 

A summary of our results is as follows:\\

\noindent \textbf{$\bullet$ Effects of the mass term (left panel of
  Fig.~\ref{fig:scalarized_sols}):}
The main effect of the mass term is to alter the threshold for the onset of scalarization, as already noted in~\cite{Brihaye:2018grv}. The dimensionless mass threshold $\hat M$, below which
scalarization is possible, is studied in Sec.~\ref{sec:radschw}, and it is plotted as a function of $\hat{\mu}$ in the
left panel of Fig.~\ref{fig:scalarized_sols} for solutions where the
radial profile of the scalar field has no nodes (black, solid line)
and one node (red, dashed line). Only the nodeless solutions are radially stable. Note that $\hat M$ is a decreasing function of
$\hat\mu$. This can be qualitatively understood from the effective potential \eqref{eq:effpot}: $\hat m_{\rm eff}$ grows
with $\hat \mu$ (at fixed $\mathcal{G}$), so that the tachyonic instability responsible for scalarization is harder to
realize.\\

\noindent \textbf{$\bullet$ Effects of the quartic self-interaction
  (right panel of Fig.~\ref{fig:scalarized_sols}):}
The quartic self-interaction stabilizes scalarized BH solutions with respect to radial perturbations. At fixed
$\hat\mu$, all scalarized solutions with $\hat\lambda<\hat\lambda_{\rm crit}$ are unstable, and we conjecture that
gravitational collapse will generally lead to a Schwarzschild solution since these are always stable for $
\hat M>\hat M_t$. This corresponds to the region on the right of the dotted vertical
lines in Fig.~\ref{fig:scalarized} below. When $\hat\lambda>\hat\lambda_{\rm crit}$, stable scalarized BH solutions are
possible: these are the solid lines on the left of the dotted vertical lines in Fig.~\ref{fig:scalarized}, and we
conjecture that they are the end-state of gravitational collapse. The threshold value $\hat\lambda_{\rm crit}(\hat\mu)$
is shown in the right panel of Fig.~\ref{fig:scalarized_sols}. Qualitatively, this can be understood as follows: For
scalarized solutions, the effective mass for the scalar [Eq.~\eqref{eq:effmass}] is tachyonic, at least in some region of spacetime, and therefore the scalar tends to grow from its scalarized
value. Introducing a quartic term in the effective potential \eqref{eq:effpot} bounds the effective potential from
below, so that there is a stable minimum about which the effective mass \eqref{eq:effmass} is positive and the solution
is globally stable. This is also the reason why a quartic Gauss-Bonnet coupling can stabilize the scalarized solutions
\cite{Minamitsuji:2018xde,Silva:2018qhn}, although in the Gauss-Bonnet case the coefficient of the $\varphi^4$ term is
also spacetime-dependent. 

{The existence of a global stable minimum should resolve the concerns raised in reference \cite{Anson:2019uto}, where it was shown that quantum fluctuations could trigger the tachyonic instability during inflation. In our model, the field would begin, and remain, at the global minimum for the duration of inflation and play no role in its dynamics (the field's mass would be much larger than the Hubble scale so that the field does not fluctuate). That being said, inflation occurs at energies far higher than the cut-off of the effective field theory for spontaneous scalarization ($10^{-20}$ GeV for scalarized solar mass BHs), and it is not clear that the range of validity of any current model exhibiting scalarization can be extended to the early Universe.   }\\

\noindent \textbf{$\bullet$ Mass range for scalarization and maximum
  scalar charge (Fig.~\ref{fig:MQ}):}
For any given choice of the theory parameters $(\hat\mu, \hat\lambda)$, marginally stable scalarized BH solutions
correspond to a minimum in the BH mass $\hat M$ and a maximum in the
scalar charge $\hat Q$: cf.~again
Fig.~\ref{fig:scalarized} below. (This maximum charge refers to
{stable} BHs; unstable BHs can have larger charges, but they
are unphysical.) In the left panel of Fig.~\ref{fig:MQ} we focus on nodeless solutions, and we plot: (i)
the $\hat\lambda$--independent threshold mass $\hat M=\hat M_{\rm t}(\hat \mu)$ below which scalarization is possible
(thick, gray line); (ii) the minimum dimensionless mass $\hat M_{\rm min}(\hat\mu)$, below which
both Schwarzschild and scalarized BH solutions are unstable, for selected values of $\hat\lambda$. The mass range in which stable, scalarized BH solutions can exist becomes narrower as
$\hat\mu$ increases.

Summarizing: for $\hat{M}>\hat{M}_{\rm t}$, the Schwarzschild solution
is stable, while scalarized BH solutions are unstable to radial
perturbations; for $\hat{M}_{\rm min}<\hat{M}<\hat{M}_{\rm t}$ {there is at least one stable $n=0$ scalarized BH,} 
while the Schwarzschild
solution (and the $n>0$ scalarized BH solutions) are unstable;
finally, for $\hat{M}<\hat{M}_{\rm min}$, all BH solutions are
unstable. The existence of a minimum BH mass is a common feature in
theories with scalar-Gauss-Bonnet coupling (see e.g. the cases of
Einstein-dilaton Gauss-Bonnet gravity~\cite{Kanti:1995vq,Pani:2009wy}
and of shift-symmetric Gauss-Bonnet gravity~\cite{Sotiriou:2014pfa}),
although in theories that do not exhibit scalarization (such as these)
the minimum mass is due to the inability to satisfy a regularity
condition at the horizon.

The right panel of Fig.~\ref{fig:MQ} shows the maximum dimensionless scalar charge $\hat Q_{\rm max}(\hat{\mu})$ for selected values of
$\hat\lambda$. The most relevant feature here is that, for all values of $\hat\lambda$ that we investigated, $\hat Q_{\rm
  max}(\hat \mu)$ has a local maximum $\sim 0.15$: this near-universal maximum value of the scalar charge is of
phenomenological interest, because the dipolar radiation in BH binaries (which is potentially measurable by
gravitational-wave interferometers) is proportional to the difference
between the BH charges~\cite{Yagi:2011xp,Berti:2015itd,Berti:2018cxi}.

\subsection{Plan of the paper}

The paper is organized as follows.
In section \ref{sec:theory} we present the equations of motion resulting from the action \eqref{eq:action} and analyze their properties.
In section \ref{sec:radschw} we investigate the effect of a nonzero scalar mass on the threshold for the onset of scalarization. We accomplish this by
studying the limit in which the scalar is decoupled from the metric equations of motion, i.e.~we consider the linearized
field equations for a scalar field propagating on a Schwarzschild background.
In section \ref{sec:bhs_sol} we move beyond this ``decoupling limit'' and solve the coupled metric-scalar equations
numerically in order to confirm the results of our linear analysis. {We also calculate the properties of the scalarized
solutions, including their stability to radial perturbations. The study of radial perturbations in scalar-Gauss-Bonnet theories is by now standard
(cf.~\cite{Blazquez-Salcedo:2018jnn,Minamitsuji:2018xde,Silva:2018qhn}), so we do not rederive the formalism in this
work.}. In section
\ref{sec:conclusion} we summarize our results and discuss possible directions for future work.

\section{Field Equations and Scalarized Solutions}
\label{sec:theory}

The modified Einstein equations can be obtained by extremizing the action \eqref{eq:action} with respect to the metric
and the scalar field, with the result
\begin{align}
G_{ab}&=T_{ab}^{\varphi}-\frac{1}{2}{\cal K}_{ab},
\label{eq:einstein_eq}
\\
\Box\varphi&=V_{,\varphi} - f_{,\varphi}{\cal G},
\label{eq:scalar_eq}
\end{align}
where
\begin{align}
  T_{ab}^\varphi&=\frac{1}{2}\partial_a\varphi\partial_b\varphi-\frac{1}{2}g_{ab}\left[\frac{1}{2}
    (\partial_c\varphi)^2+V(\varphi)\right],\\
	{\cal K}_{ab}&=2g_{c(a}g_{b)d}\epsilon^{edjg}\nabla_h\left[^*{R^{ch}}_{jg}f'\nabla_e\varphi\right],
\end{align}
where $V(\varphi)$ is given in equation \eqref{eq:potential}, 
$f(\varphi)=(\eta\varphi^2+\zeta\varphi^4)/8$, and $^*R^{ab}_{cd}=\epsilon^{abef}R_{efcd}$.

As discussed in~\cite{Doneva:2017bvd,Silva:2017uqg}, Schwarzschild
solutions exist in scalar Gauss-Bonnet theories provided that there is
some $\bar\varphi$ such that $f'(\bar\varphi)=0$. In our model,
$\bar\varphi=0$. Allowing for a nonzero value of the background scalar
may have important phenomenological consequences for gravitational
wave astronomy, as pointed out in the context of
Einstein-Maxwell-dilaton theory~\cite{Julie:2017rpw}, and we plan to
revisit this assumption in future work.

We focus on static, spherically symmetric BHs. In this case the line element and the scalar field read
\begin{align}
\dd s^2&=-A (r) \dd t^2+B(r)^{-1}\dd r^2+r^2\dd\Omega^2,\label{eq:metric}\\
\varphi&=\varphi_0(r),
\label{eq:scalar_f}
\end{align}
where $\dd\Omega = \dd \theta^2 + \sin^2\theta\, \dd \phi^2$ is the line element on a 2-sphere. The field equations can be obtained by substituting
Eqs.~\eqref{eq:metric} and \eqref{eq:scalar_f} into Eqs.~\eqref{eq:einstein_eq} and~\eqref{eq:scalar_eq}.
We show the equations in Appendix~\ref{app:equations}. We also make them available online through a {\sc Mathematica} notebook~\cite{notebook}.

The field equations must be supplemented by boundary conditions. Spherically symmetric BHs have an event horizon $r_h$
where the functions $A$ and $B$ vanish and the scalar field tends to a constant:
\begin{align}
	A(r\approx r_h) &\approx a_1(r-r_h)+{\cal O}[(r-r_h)^2],\label{eq:metric_cond}\\
	B(r\approx r_h) &\approx b_1(r-r_h)+{\cal O}[(r-r_h)^2],\\
	\varphi_0(r\approx r_h)&\approx \varphi_{0h}+{\cal O}[(r-r_h)].
\end{align}
These conditions impose a restriction on the derivative of the scalar field at the horizon:
\begin{equation}
\left.\frac{\dd \varphi_{0}}{\dd r}\right|_{r=r_{\rm h}}=a^{-1}\left(b+c\sqrt{\Delta}\right),
\label{eq:phi_der}
\end{equation}
where $a,\,b,$ and $c$ are functions of $r_h$, $\varphi_{0h}$, and of
the parameters of the theory. {The explicit expression
  of these functions is given in Appendix~\ref{app:equations}}.
The important quantity is $\Delta$, which is given by
\begin{align}
  \Delta&=1-6\frac{\varphi_{0h}}{r_h^4}\left(\eta+\zeta\varphi_{0h}^2\right)^2\Bigg\{1-\nonumber\\
  &\quad-\frac{1}{2}\varphi_{0h}^2\left(\eta+\zeta\varphi_{0h}^2\right)\left(\mu^2+2\lambda\varphi_{0h}^2\right)\nonumber\\
  &\quad-\frac{r_h^2}{6}\varphi_{0h}^2\left(\mu^2+\lambda\varphi_{0h}^2\right)\left[1
    +\frac{1}{16r_h^2}\left(\eta\varphi_{0h}+\zeta\varphi_{0h}^3\right)^2\times\right.\nonumber\\
    &\quad\left.\left.\times\left(-\frac{24}{r_h^2}+\mu^2\varphi_{0h}^2
    +\lambda\varphi_{0h}^4\right)\right]\right\}\,.\label{eq:deltaexpr}
\end{align}
When $\Delta<0$ it is not possible to enforce regularity at the horizon. Ref.~\cite{Sotiriou:2014pfa} studied this
regularity condition for shift-symmetric scalar-Gauss-Bonnet gravity, showing that there is a naked singularity
when the condition is violated. Thus, $\Delta>0$ is a necessary condition for the existence of BH solutions.

By expanding the field equations for large $r$ we obtain
\begin{align}
    A(r\gg r_{\rm h})&\simeq 1-{2M}/{r}\,,
  \label{eq:metric_inf}\\
  B(r\gg r_{\rm h})&\simeq 1-{2M}/{r}\,,
    \\
  \varphi_0(r\gg r_{\rm h})&\simeq {Q\,e^{-\mu r}}/{r}\label{eq:scalar_inf}\,,
\end{align}
where $M$ is the ADM mass, $Q$ is an integration constant, and we have set the cosmological value of the scalar field to
zero.
In the $\mu\to0$ limit, the scalar field decays like $\varphi\sim 1/r$, and the constant in front of $1/r$ is typically referred to
as the ``scalar charge.'' Strictly speaking, $Q$ is not a conserved
charge (even when $\mu=0$), but we will follow conventions and refer to it as such from here on.

Typically, in scalar-tensor theories one must set the scalar field's mass such that the force range is sub-micron (for
$\mathcal{O}(1)$ couplings), or else the theory will fail laboratory and solar system tests of
GR~\cite{Adelberger:2003zx,Adelberger:2006dh,Adelberger:2009zz,Antoniadis:2011zza,Sakstein:2015zoa,Sakstein:2015aac,Burrage:2016bwy,Burrage:2017qrf,Sakstein:2017pqi}.
Therefore one would expect the spacetime outside the BH to rapidly approach the Schwarzchild metric, thereby suppressing
any deviations from GR. This logic follows from scalar-gravity couplings of the form $\varphi R$, which, in the absence
of any screening mechanisms, give rise to Yukawa forces. The coupling considered in our model $\eta\varphi^2\mathcal{G}$
is expected to appear at high post-Newtonian order in the weak-field limit (provided $\eta/M_\odot^2$ is not too large),
and therefore the theory is compatible with Solar System tests of GR \cite{EspositoFarese:2004cc,Sotiriou:2006pq}. Furthermore, since it is unlikely that weakly
gravitating objects like the Sun and the Earth are scalarized, gravity in the Solar System should behave identically to
GR. For these reasons, we will not place any restrictions on the mass of the field in this work. One could
  imagine completing the EFT by adding a term proportional to $\varphi^2R$ into the action, which is not forbidden by the
  symmetries, and which we have ignored in this work for the sake of simplicity. Such couplings could give rise to
  Yukawa-like forces, but (again) only if the Sun or the Earth is scalarized, which is unlikely to be the case, with the
  exception of extreme couplings~\cite{Sakstein:2017lfm,Sakstein:2017nns}. The situation would be different if the asymptotic field value were different from zero.

\section{Schwarzschild radial stability and the scalarization threshold}
\label{sec:radschw}

In Sec.~\ref{sec:bhs_sol} we will explore the BH solutions of the theory. Before doing so, we first
wish to understand whether such solutions can exist as a result of instabilities of the ordinary Schwarzschild solution to linear perturbations.  

The Schwarzschild metric with a vanishing scalar field is a solution of Eqs.~\eqref{eq:einstein_eq}
and~\eqref{eq:scalar_eq}.  We can study the radial stability of the Schwarzschild spacetime by considering perturbations
of the field equations of the form
%
$\varphi=\varepsilon \varphi_1(r) e^{-i\omega t}/r$,
%
where $\varepsilon$ is a small bookkeeping parameter. From the scalar field equation~\eqref{eq:scalar_eq} we find
\begin{equation}
\frac{\dd^2 \varphi_1}{\dd r_\ast^2}+(\omega^2-V_{\rm eff})\varphi_1=0\,,
\label{eq:pert_sch}
\end{equation}
where
\begin{equation}
V_{\rm eff}=\left(1-\frac{r_h}{r}\right)\left(\frac{r_h}{r^3}+\mu^2-\frac{3\, \eta\, r_h^6}{r^6}\right)\,.
\label{eq:pot_sch}
\end{equation}
This equation involves only the field's mass $\mu$ and the strength of the Gauss-Bonnet coupling $\eta$. Therefore,
higher-order terms in the scalar potential and in the coupling function do not have any influence on the stability of
the Schwarzschild spacetime. In particular, the threshold for scalarization is independent of $\lambda$.

\begin{figure}
\includegraphics[width=\columnwidth]{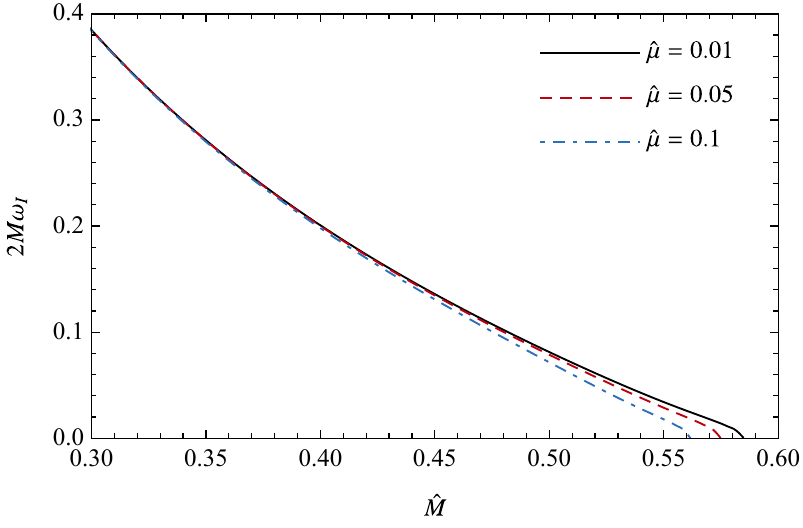}
\caption{Frequency of the unstable modes of Schwarzschild BHs in our theory.  The frequency becomes zero at $\hat M=\hat M_t$.}
\label{fig:sch_ins}
\end{figure}

\begin{figure}
\includegraphics[width=\columnwidth]{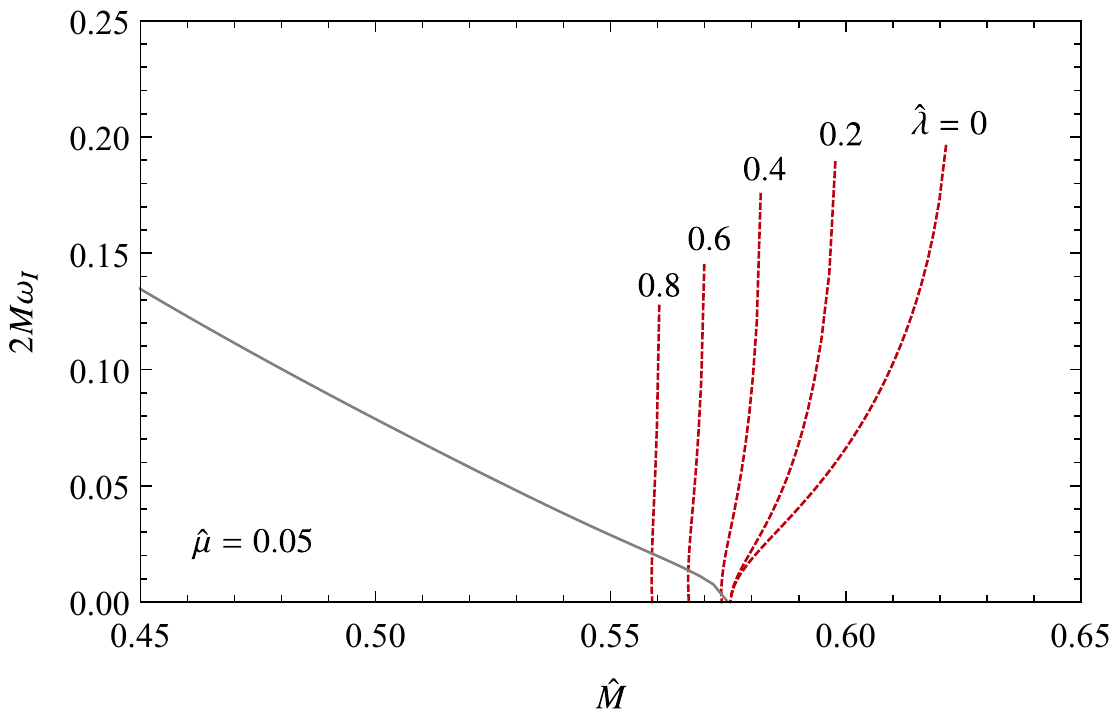}
\caption{Unstable modes of scalarized BHs, compared with the Schwarzschild case (gray solid line), for the representative case $\hat \mu=0.05$.}
\label{fig:radialstab}
\end{figure}

To investigate the radial stability of the Schwarzschild spacetime, we solve Eq.~\eqref{eq:pert_sch} by requiring that
the field vanishes at the BH horizon and at infinity~\cite{Silva:2018qhn}.  Since the equation is real, the eigenvalue
$\omega^2$ is also real, and $\omega^2<0$ corresponds to unstable
modes~\cite{Kanti:1997br,Blazquez-Salcedo:2018jnn,Silva:2018qhn}.  The critical threshold value $\hat{M}=\hat{M}_{\rm t}$ for
which scalarization can occur corresponds to solutions of Eq.~\eqref{eq:pert_sch} with eigenvalue $\omega=0$, indicating
a transition between stable and unstable states. The condition $\omega^2=0$ is satisfied by different values of
$\hat{M}$, corresponding to scalarized solutions with $n=0,1,\dots$
nodes in the scalar field profile. We denote the
threshold value for the $n=0$ solution by $\hat{M}_{\rm t}$.

In the left panel of Fig.~\ref{fig:scalarized_sols} we show
$\hat{M}_{\rm t}$ as a function of the mass $\hat\mu$ of the field (we
also show the threshold values for the $n=1$ solution, which, as
discussed in the introduction, is always smaller than the threshold
mass $\hat{M}_{\rm t}$ for $n=0$). {One can see that the
threshold for scalarization $\hat{M}_{\rm t}$ decreases with increasing $\hat\mu$. This can be understood by considering the effective mass for the scalar given in Eq.~\eqref{eq:effmass}:
larger values of $\mu$ require the product $\eta\mathcal{G}$ to be larger in order to induce the tachyonic instability. The instability is therefore harder to realize for larger scalar masses. }

By solving Eq.~\eqref{eq:pert_sch} we can also investigate the instability time scale as a function of $\hat{M}$. In
Fig.~\ref{fig:sch_ins} we show the normalized frequency for unstable modes, $2M\omega_I$, as a function of the parameter
$\hat{M}$. The Schwarzschild solution is stable ($\omega_I<0$) in the region $\hat{M}>\hat{M}_{\rm t}$, where
$\hat{M}_{\rm t}$ is the value corresponding to the intersection of $2M\omega_I$ with the x-axis of this plot. The three
cases studied here correspond to the blue dots in the left panel of Fig.~\ref{fig:scalarized_sols}. It is therefore plausible that hairy solutions should exist in the region $\hat{M}<\hat{M}_{\rm t}$, where the Schwarzschild BH is
unstable. This expectation will be confirmed in Sec.~\ref{sec:bhs_sol} below.

\begin{figure*}%
\includegraphics[width=0.7\columnwidth]{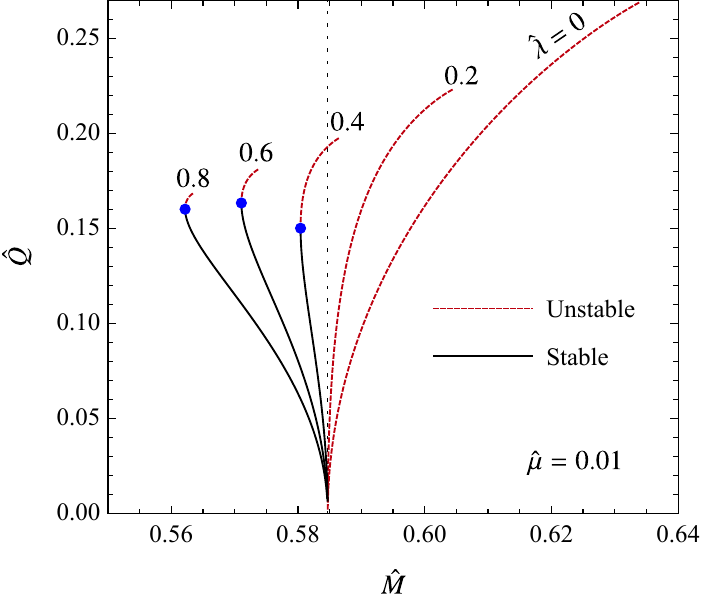}\includegraphics[width=0.7\columnwidth]{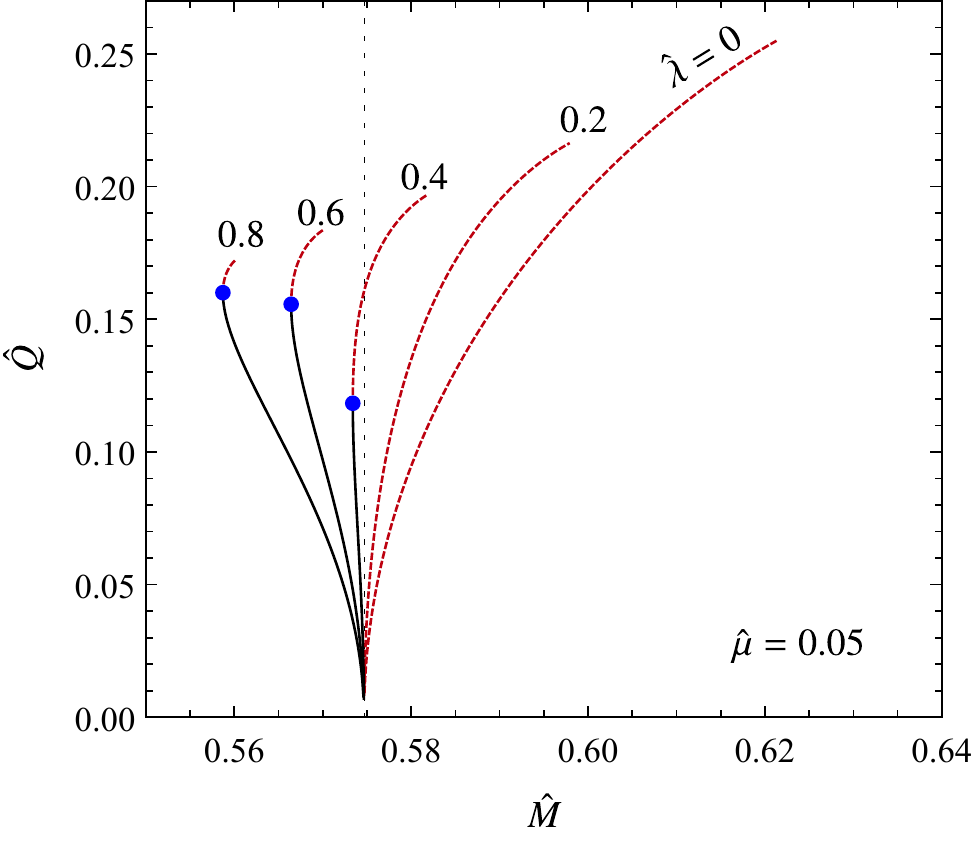}\includegraphics[width=0.7\columnwidth]{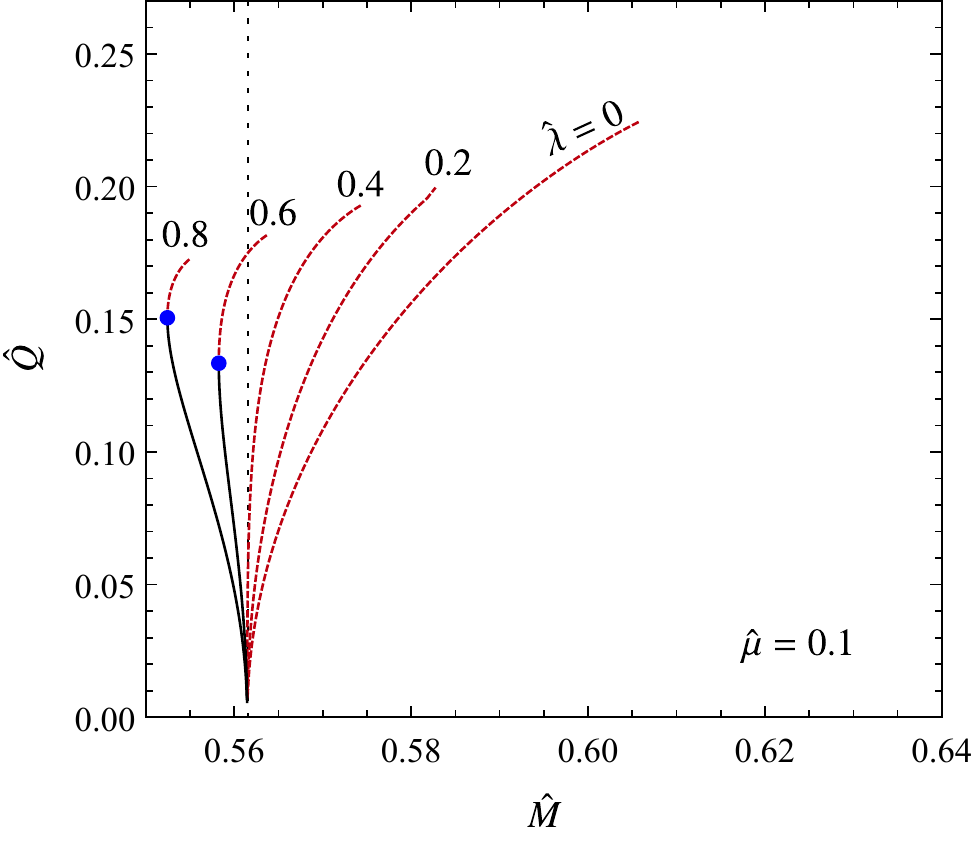}
\caption{Charge-mass diagram for scalarized solutions with a quadratic scalar-Gauss-Bonnet coupling ($\zeta=0$) and the scalar
  potential of Eq.~\eqref{eq:potential}. The threshold mass $\hat{M}_{\rm t}$ corresponds to the dotted vertical line.
  For $\hat{M}>\hat{M}_{\rm t}$ scalarized solutions are radially unstable, while the Schwarzschild solution is stable. When
  $\hat\lambda$ is large enough, we can have solutions with $\hat{M}<\hat{M}_{\rm t}$. In this region there are two branches
  of scalarized BH solutions: the upper branch (dashed lines) is unstable to radial perturbations, whereas the lower
  branch (solid lines) is stable. Blue dots mark solutions with marginal stability, which correspond to the minimum
  mass, maximally charged scalarized BH for the given $(\hat\mu,\,\hat\lambda)$.}
\label{fig:scalarized}
\end{figure*}

\section{Scalarized black hole solutions and radial stability}
\label{sec:bhs_sol}

In this section we solve the fully nonlinear equations to construct scalarized solutions, and check their stability under linear radial scalar and tensor perturbations. This is accomplished as follows. First, we integrate the field equations outwards starting from the horizon, where we impose the conditions
\eqref{eq:metric_cond}--\eqref{eq:phi_der}. By matching the numerical solutions with
Eqs.~\eqref{eq:metric_inf}--\eqref{eq:scalar_inf} in the far region ($r\gg r_h$), we can extract the BH mass $\hat{M}$
and the scalar charge $\hat{Q}$. This procedure gives us the unperturbed solution. Next we check stability. The linearized field equations for radial perturbations follow from
the ansatz
\begin{align}
\varphi&=\varphi_0+\varepsilon\frac{\varphi_1}{r},\label{eq:field_pert}\\
\dd s^2&=[A+\varepsilon F_t(t,r)]\dd t^2+[B^{-1}+\varepsilon F_r(t,r)]+r^2\dd \Omega^2,\label{eq:metric_pert}
\end{align}
where $(A,\,B,\,\varphi_0)$ are functions of $r$ which satisfy the
zeroth-order (background) field equations. By inserting
Eqs.~\eqref{eq:field_pert} and~\eqref{eq:metric_pert} into the field
equations~\eqref{eq:einstein_eq} and~\eqref{eq:scalar_eq} and
expanding up to first order, one can show that the equations for the
perturbation functions reduce to a single second-order equation of the
form
\begin{equation}
  h(r)\frac{\partial^2\varphi_1}{\partial t^2}-\frac{\partial^2\varphi_1}{\partial r^2}+k(r)
  \frac{\partial\varphi_1}{\partial r}+p(r)\varphi_1=0,
\label{eq:per_scalar}
\end{equation}
(see Appendix~\ref{app:equations} and the supplemental {\sc
  Mathematica} notebook~\cite{notebook}) where the coefficients
$(h,\,k,\,p)$ depend only on the background quantities and on $r$ (cf.~\cite{Kanti:1995vq,Blazquez-Salcedo:2018jnn,Silva:2018qhn}).
Eq.~\eqref{eq:per_scalar} can be further manipulated to reduce it to a
Schr\"odinger-like form, but since this step is not necessary to
analyze the stability of the system, and generates more
complicated coefficients, we prefer not to display it here
(see~\cite{Blazquez-Salcedo:2018jnn} for details). A mode analysis can
be performed by looking for solutions of the form
$\varphi_1(t,r)=\varphi_1(r)e^{-i\omega t}$, and by imposing the
requirement that $\varphi_1(r)$ vanishes at the horizon and at
infinity when searching for unstable modes. These requirements (as in
Sec.~\ref{sec:radschw}) result in an eigenvalue problem for
$\omega^2<0$.

Before applying this process in general, it is instructive to perform a preliminary comparative study in order to discern how self-interactions affect the stability of scalarized solutions. In Fig.~\ref{fig:radialstab} we fix $\hat\mu=0.05$ and we compare the
normalized imaginary mode for the scalarized solutions with the
corresponding calculation for the Schwarzschild case, as presented in
Fig.~\ref{fig:sch_ins}.  When $\hat\lambda\leq 0.2$, both the modes of
the scalarized solutions (dashed red) and the Schwarzschild modes
(solid gray) converge to zero when $\hat M=\hat M_t$. However, for
$\hat \lambda >0.2$ the modes tend to zero when
$\hat M=\hat M_{\rm min}$ and $\hat Q=\hat Q_{\rm max}$, and we found
no unstable modes for BHs with $\hat M>\hat M_{\min}$ and
$\hat Q<\hat Q_{\rm max}$. We note also that the unstable mode
frequencies typically decrease as $\hat\lambda$ increases, {implying stability on longer time-scales}.
Qualitatively similar conclusions apply to other values of $\hat\mu$.

The main results of our integrations are presented in
Fig.~\ref{fig:scalarized}, where we show scalarized solutions in the
$(\hat{M},\,\hat{Q})$ plane for representative values of $\hat\mu$ and
$\hat\lambda$. The dotted vertical line represents the threshold for
the stability of the Schwarzschild solution,
$\hat{M}=\hat{M}_{\rm t}$. Solid lines correspond to radially stable
solutions, while dashed lines correspond to radially unstable
solutions. Note that we use different conventions for radial stability
with respect to Refs.~\cite{Blazquez-Salcedo:2018jnn,Silva:2018qhn},
where solid and dashed lines have the opposite meaning.

When $\hat{\lambda}=0$, all scalarized solutions are in the region $\hat{M}>\hat{M}_{\rm t}$,
where the Schwarzschild solution is stable. These scalarized solutions are radially unstable, and it is
plausible that Schwarzschild BHs will be the end-state of gravitational collapse. As $\hat{\lambda}$ increases, the solutions move into the region where $\hat{M}_{\rm min}<\hat{M}<\hat{M}_{\rm t}$; the
minimum mass $\hat{M}_{\rm min}$ corresponds to the blue dots in Fig.~\ref{fig:scalarized}. Schwarzschild BHs are
unstable in this region, so the BH can support a nontrivial scalar provided the scalarized solutions are stable.  For
${\hat M}<\hat{M}_{\rm min}$, both Schwarzschild and scalarized BHs are unstable.

Our analysis reveals that the quartic self-interaction can stabilize scalarized solutions with a quadratic
scalar-Gauss-Bonnet coupling up to some maximum scalar charge $\hat Q$, beyond which the solutions are
unstable. Interestingly, it is possible to have two scalarized solutions (in addition to the unstable Schwarzschild
solution) at fixed $\hat M$, provided that $\hat\lambda$ is large enough. In such cases, the solution with larger $\hat
Q$ is unstable, and is expected to decay to the solution with smaller $\hat Q$, which is stable.

The main result of this section is that we do not need more exotic scalar-Gauss-Bonnet
couplings to stabilize the scalarized solutions: leading-order scalar self-interactions are sufficient. From an EFT
perspective, these models are better-motivated.

\section{Discussion and Conclusions}
\label{sec:conclusion}

Black hole spontaneous scalarization is so far the only known mechanism that allows BHs to possess scalar hair only if their mass is below a certain threshold. Theories that allow for this phenomenon are prime candidates for modelling deviations form GR that have so far avoided detection but can be tested using current and future gravitational wave observations. It therefore behooves the theoretical community to devise robust, stable theories that exhibit BH scalarization. To date, all studies in the literature are not consistent effective field theories since they ignored leading-order terms that are compatible with the underlying symmetries of the theory. The aim of the present work is to take the first steps towards realizing the phenomenon within robust and well-motivated theories.

In this paper, we have presented the simplest model that exhibits spontaneous scalarization by viewing the theory as one
of a $\mathbb{Z}_2$-symmetric scalar and writing down all of the leading-order (relevant and marginal) operators, as
well as the leading-order coupling of the scalar to the Gauss-Bonnet
invariant required to produce scalarized BHs. In
practice, this is tantamount to including a mass and quartic self-interaction for the scalar, so that the theory
includes a massive scalar with a $\phi^4$-potential and a quadratic coupling of the scalar to the Gauss-Bonnet
invariant.

Our analysis has revealed that spontaneous scalarization persists in
this bottom-up construction. We have demonstrated that static
scalarized solutions exist and, furthermore, that they are stable to
radial perturbations. This model (possibly augmented by a $\varphi^2R$ coupling, and other
dimension-six operators) therefore represents the leading candidate
model with which to explore spontaneous scalarization. In future
studies, we intend to take this program forward by studying rotating
BHs and neutron stars, and by understanding the stability and dynamics
of these compact objects in full generality. The ultimate aim of this
program is to predict theoretically sound observational
signatures that can be used to test GR in the strong-field regime with
upcoming gravitational wave observations.\\

\section*{Acknowledgments}
This work was supported by the H2020-MSCA-RISE-2015 Grant No.  StronGrHEP-690904 and by the COST action CA16104
``GWverse''.
H.O.S was supported by NASA grants NNX16AB98G and 80NSSC17M0041.
J.S. was supported by funds provided to the Center for Particle Cosmology by the University of Pennsylvania.
E.B. is supported by NSF Grants No. PHY-1841464, AST-1841358, PHY-090003, and NASA ATP
Grant No. 17-ATP17-0225.
C.F.B.M. would like to thank Conselho Nacional de Desenvolvimento Científico e Tecnológico (CNPq) for partial financial support, 
and the Johns Hopkins University for kind hospitality during the preparation of this work and
the American Physical Society which funded the visit through the International Research Travel Award Program.\\

\appendix
\section{Spherical black holes in scalar Gauss-Bonnet gravity}\label{app:equations}

From the nontrivial components of the zeroth-order Einstein equations, we obtain
\begin{widetext}
\begin{align}
(t,t): \qquad &B \left\{ \left(\varphi_{0}''\right)^2 \left[16 (B-1) f_{,\varphi_{0}\varphi_{0}} -r^2\right]+16
        (B-1) \varphi_{0}'' f_{,\varphi_{0}} -4\right\}
         -4 B' \left[2 (1-3 B) \varphi_{0}' f_{,\varphi_{0}}+r\right]-r^2 V(\varphi_{0} )+4=0, \\
(r,r): \qquad &\frac{1}{4} A \left\{ B \left[4-r^2 \left(\varphi_{0}''\right)^2\right]+r^2 V(\varphi_{0}
    )-4\right\}+B A' \left[2 (1-3 B) \varphi_{0}' f_{,\varphi_{0}} + r\right]=0, \\
   (\theta,\theta): \qquad &-r B A'' \left(r-4 B \varphi_{0}' f_{,\varphi_{0}}\right)+4 r B^2
   A' \varphi_{0}'' f_{,\varphi_{0}}
   -\frac{1}{2} r^2 A \left[B
   \varphi_{0}'^2+V(\varphi_{0} )\right] \nonumber\\
    & + \frac{1}{2A}\left[r B A'^{2}
    \left(r-4 B \varphi_{0}' f_{,\varphi_{0}} \right)\right] + A' \left\{r B
   \left[4 B \varphi_{0}'^2 f_{,\varphi_{0}\varphi_{0}} - 1\right]-\frac{1}{2} r
   B' \left(r-12 B \varphi_{0}' f_{,\varphi_{0}} \right)\right\}-r A B'=0 \,,
\end{align}
where a prime indicates differentiation with respect to $r$. The equation for the background scalar field is
\begin{align}
&A B \varphi_{0}'' +\frac{1}{2} \varphi_{0}'
    \left[BA' +A \left(\frac{4 B}{r}+B'\right)\right]+\frac{4}{r^2}\left[(B-1)
    B  A'' f_{,\varphi_{0}}  \right]\nonumber \\
& -\frac{1}{2 r^2 A} \left\{4 A' \left[(B-1)
    B A' + A (1-3 B) B'\right] f_{,\varphi_{0}}+r^2 A^2 V_{,\varphi_{0}}\right\}=0.
\label{eq:scalar_back}
\end{align}
\end{widetext}

The equations above can be recast as a system of three differential
equations: two first-order equations for $A$ and $B$, and
one second-order equation for $\varphi_0$. These are integrated as explained in Sec.~\ref{sec:theory}, in units where $r_h=1$, changing the parameters $(\eta,\zeta,\lambda,\mu)$ in each integration. As noted in
Ref.~\cite{Silva:2017uqg}, only some values of the parameters allow for scalarized solutions. With these we
construct the background solutions shown in Fig.~\ref{fig:scalarized}.

Specializing to the quartic coupling and the quartic potential, the coefficients
appearing in the condition on the scalar field derivative at the
horizon of Eq.~\eqref{eq:phi_der} are given by
\begin{align}
a&=\frac{\varphi_{0h}}{r_h}\op(\eta+\zeta \varphi_{0h}^2\cl)\left\{-4
+2 \varphi_{0h}^2 \left(\zeta  \varphi_{0h}^2+\eta \right) \left(2 \lambda  \varphi_{0h}^2+\mu ^2\right) \right.
\nonumber \\
&\quad \left.+\frac{\varphi_{0h}^2}{r_h^2} \left(\lambda  \varphi_{0h}^2+\mu ^2\right)
\left[r_h^4-\varphi_{0h}^2 \left(\zeta  \varphi_{0h}^2+\eta \right)^2\right]\right\},\\
b&=\varphi_{0h}^2 r_h^2 \left(\lambda  \varphi_{0h}^2+\mu ^2\right)-4,\\
c&=4-\varphi_{0h}^2 r_h^2 \left(\lambda  \varphi_{0h}^2+\mu ^2\right)+\frac{3 \varphi_{0h}^4}{r_h^2}
\left(\zeta  \varphi_{0h}^2+\eta \right)^2 \left(\lambda  \varphi_{0h}^2+\mu ^2\right)
\nonumber \\
&\quad -\frac{1}{4} \varphi_{0h}^6 \left(\zeta  \varphi_{0h}^2+\eta \right)^2 \left(\lambda  \varphi_{0h}^2+\mu ^2\right)^2
\nonumber \\
&\quad -4 \varphi_{0h}^2 \left(\zeta  \varphi_{0h}^2+\eta \right) \left(2 \lambda  \varphi_{0h}^2+\mu ^2\right).
\end{align}

The equations describing the perturbations can be obtained by
expanding the Einstein-scalar system up to first order.  The nonzero
components of the perturbed Einstein equations are $(t,t)$, $(t,r)$,
$(r,r)$, and $(\theta,\theta)$. Additionally, we have one more
equation from the first-order expansion of the scalar field
equation. These five equations can be manipulated to
obtain~\eqref{eq:per_scalar} with a procedure similar to the one
presented in~\cite{Kanti:1997br,Blazquez-Salcedo:2018jnn}. Instead of
showing the explicit form of the differential equations, which
are rather lengthy, we provide a companion {\sc Mathematica} notebook
which shows the nontrivial components of the first-order Einstein
equations and the procedure to obtain Eq.~\eqref{eq:per_scalar} from
these equations~\cite{notebook}.\\

\begin{figure}[ht]
\includegraphics[width=\columnwidth]{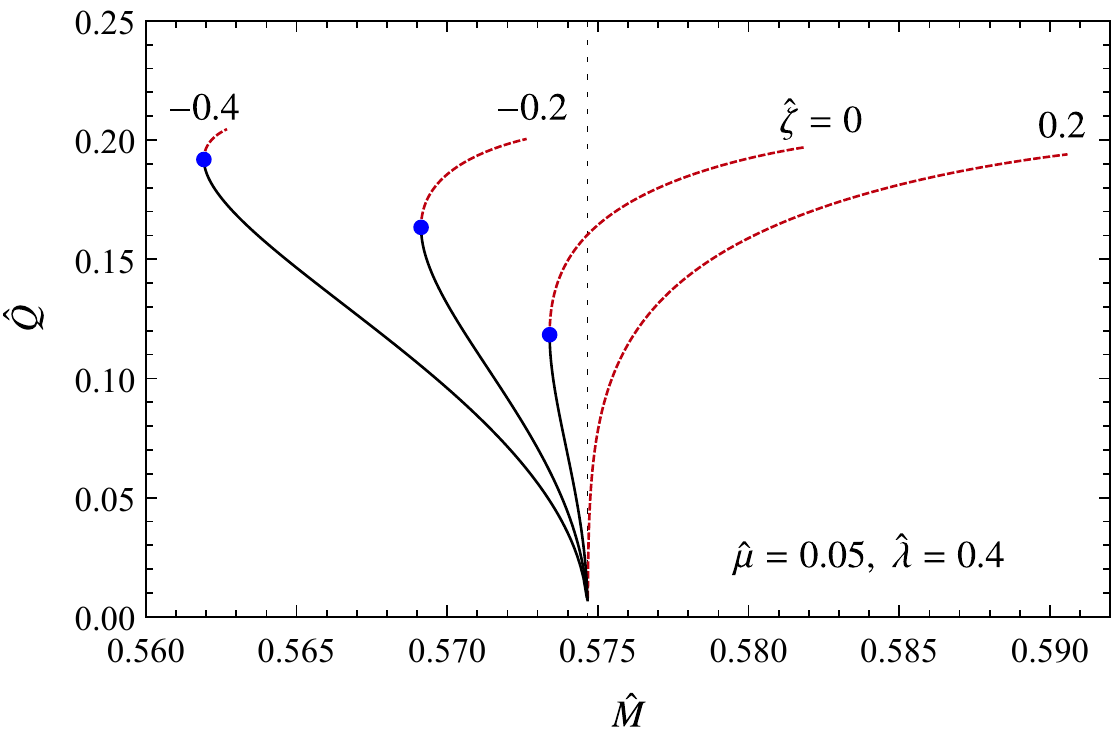}
\caption{Charge-mass diagram considering a quartic term in the
  coupling function and a quartic self-interacting potential.}
\label{fig:quartic}
\end{figure}

\section{Self-interactions within quartic Gauss-Bonnet coupling}\label{app:quartic}

Ref.~\cite{Silva:2018qhn} showed that scalar-Gauss-Bonnet theories with
$V=0$ and a quartic coupling term
\begin{equation}
  f(\varphi)=\frac{1}{8}\left(\eta\varphi^2+\zeta\varphi^4\right)
\end{equation}
with $\zeta/\eta<0$ can also
generate stable BH solutions. A natural question is whether the
combined effects---the quartic potential and the quartic coupling---can work together to stabilize BHs. While this is out of the scope of
the EFT picture, with the quartic coupling being a sub-leading
operator compared with the quartic self-interaction, here we
investigate this issue as a complement to our main results.

In Fig.~\ref{fig:quartic} we show the scalarized BH solutions
considering $(\hat \mu,\hat\lambda)=(0.05,0.4)$ for different values
of $\hat\zeta$. As expected, the quartic term in the coupling still
helps to generate stable BH solutions, even when the self-interaction
potential is present. We note that this case also exhibits a minimum
mass $\hat M_{\rm min}$ and a maximum charge $\hat Q_{\rm max}$,
unlike theories with $V=0$: cf. Fig.~2 of~\cite{Silva:2018qhn}.

\bibliography{biblio}

\end{document}